\documentstyle[aps,pre,epsf,floats]{revtex}
\begin{document}

\twocolumn[\hsize\textwidth\columnwidth\hsize\csname@twocolumnfalse%
\endcsname

\draft

\title{Chaotic behavior and damage spreading in the Glauber Ising model - 
a master equation approach }

\author{Thomas Vojta}
\address{Institut f\"ur Physik, Technische Universit\"at, D-09107 Chemnitz, 
Germany and \\ Department of Physics and Materials Science Institute, 
University of Oregon, Eugene, OR97403, USA}
\date{\today}
\maketitle

\begin{abstract}
We investigate the sensitivity of  the time evolution of  a kinetic Ising model 
with Glauber dynamics against the initial conditions.
To do so we apply the "damage spreading" method,
i.e., we study
the simultaneous evolution of two identical systems subjected to the
same thermal noise. We derive a master equation for the joint 
probability distribution of the two systems. 
We then solve this 
master equation within an effective-field approximation which goes beyond
the usual mean-field approximation by retaining the fluctuations though in a 
quite simplistic manner.
The resulting effective-field theory is then applied to different physical
situations. It is used to analyze the fixed points of the master
equation and their stability and to identify regular and chaotic phases of the Glauber
Ising model.  We also discuss the relation of our results to directed 
percolation.
\end{abstract}

]  

\section{Introduction}
The physics of  dynamic phase transitions and dynamic critical phenomena has been 
a subject of great interest for the last two decades. Whereas the dynamic behavior at
and close to usual static phase transitions is well understood \cite{hoha,langer} much 
less is known about dynamic phase transitions which do not have a static counterpart.
Sometimes it is not even known whether or not a particular dynamic transition coincides
with an equilibrium phase transition.

One of these dynamic phenomena is the so-called "damage spreading"
\cite{kauffman,derrida1,stanley}.
The central question of this problem is how a small perturbation (called the damage)
in a cooperative system changes during the further time evolution. 
Among the simplest of such cooperative systems are kinetic Ising models where
the above question has been investigated by means of Monte-Carlo simulations 
\cite{derrida1,stanley}. In these simulations two identical Ising models with different 
initial conditions 
are subjected to the same thermal noise, i.e. the same random numbers 
are used in the Monte-Carlo procedure. 
In analogy to the physics of chaotic dynamics \cite{schuster} the differences 
in the microscopic configurations of the two systems are then used to characterize the 
dynamics and to distinguish regular and chaotic phases, depending on
external parameters (e.g. temperature, magnetic field).

Later the name "damage spreading" has also been applied to a different
though related type of investigations in which the two systems are {\em not}
identical. Instead, one or several spins in 
one of the copies are permanently fixed in one direction. Therefore the
equilibrium properties of the two systems are different and
the microscopic differences between the two copies can be related to
static and dynamic correlation functions \cite{coniglio,glotzer}. 
Note that in this type
of simulations it is not essential to use identical noise (i.e. random numbers) for the
two systems. Instead it is only a convenient method to reduce the statistical error.

Whereas this second type of damage spreading is well understood and 
established as a method to numerically calculate equilibrium properties,
much less is known about the original problem of damage spreading,
viz. how sensitive is the dynamics of the Ising model to different {\em initial conditions}.
In particular, there are no rigorous results on the transition 
between regular and chaotic behavior (called the "spreading transition").

There are two different mechanisms by which the damage can spread
in a kinetic Ising model. First, the damage can spread {\em within} a single
ergodic component (i.e. a pure state or free energy valley)
of the system. This is the case
for Glauber or Metropolis dynamics. Numerical
simulations here consistently give a
transition temperature slightly lower than the equilibrium 
critical temperature \cite{ts}. 
Grassberger \cite{conjecture} conjectured that the spreading transition
falls into the universality class of directed percolation if it does not coincide
with another phase transition. This was supported by high-precision numerical
simulations for the Glauber Ising model \cite{grassberger}. 

Second, the damage can spread when the system
selects one of the free energy valleys at random after a quench from high
temperatures to below the equilibrium critical temperature. 
This is the only mechanism to produce damage spreading in an Ising
model with heat-bath dynamics.  In this case the spreading temperature 
seems to coincide with the equilibrium critical temperature below which
the two pure states separate \cite{derrida1,grassheat,heat}. Thus, at the 
spreading point there are long-range static correlations in the systems,
and the transition is expected to fall into a universality class different from
directed percolation.

In this paper we investigate the damage spreading in the Glauber Ising model by
deriving and solving a master equation for the time evolution of a joint probability 
distribution for two identical systems with different initial conditions and subjected
to the same thermal noise. The paper is organized as follows. In Section IIA we 
define the model. Transition probabilities between the states of a spin pair are
calculated in section IIB and the master equation for the joint probability 
distribution is derived in section IIC.
 We discuss how to construct a mean-field approximation for this
equation in section IIIA.  In sections IIIB, IIIC and IIID we present solutions
of the master equation within this approximation for different physical situations.
Finally, section IV is dedicated to conclusions and an outlook on future work.
A short account of part of this work has already been published \cite{vojta}
together with a comparison to the heat-bath Ising model.

\section{A master equation for damage spreading}
\subsection{The Glauber Ising model}
We consider two identical kinetic Ising models with $N$ sites described by
the Hamiltonians $H^{(1)}$ and $H^{(2)}$ given by
\begin{equation}
H^{(n)} = - \frac 1 2  \sum_{ij} J_{ij} S_i^{(n)} S_j^{(n)} - h \sum_i S_i^{(n)} 
\end{equation}
where $S_i^{(n)}$ is an Ising variable with the values $\pm 1$ and $n=1,2$ 
distinguishes the two copies. $J_{ij}$ is the exchange interaction between the
spins and $h$ denotes an external magnetic field.
The dynamics of the Ising models
is given by  the Glauber algorithm, i.e. 
in every time step a lattice site $i$ is chosen at random (the {\em same} site in both
copies). The new value of
the spin at this site is calculated according to
\begin{equation}
S_i^{(n)} (t+1) = {\rm sign} \left\{ v[h_i^{(n)}(t)] 
- \frac 1 2 +S_i^{(n)}(t) \biggl[ \xi_i(t) - \frac 1 2 \biggr] \right\} 
\end{equation}
where the transition probality $v(x)$ is given by the usual
Glauber expression
\begin{equation}
v(x) = {e^{x/T}/ ({e^{x/T}+ e^{-x/T}}}).
\end{equation}
Here $h_i^{(n)}(t)=\sum_j J_{ij} S_j^{(n)}(t) + h$ 
is the local magnetic field at site $i$ and (discretized)
time $t$ in the system $n$. $\xi_i(t) \in [0,1)$ is a random number which is identical
for both systems, and $T$ denotes the temperature. The spins at all sites 
different from site $i$
are unchanged within this time step.

The central quantity in any damage spreading process is the 
distance between the two systems in phase space, called the
Hamming distance (or the damage). It is defined by
\begin{equation}
D(t) = \frac 1 {2N} \sum_{i=1}^N |S_i^{(1)}(t) - S_i^{(2)}(t) |
\end{equation}
and identical to the portion of sites where the spins in the two systems
differ. 

In order to describe the simultaneous time evolution of the two systems
$H^{(1)}$ and $H^{(2)}$
we define a variable $\nu(t)$ at each lattice site which describes 
the state of a spin pair $(S^{(1)},S^{(2)})$. 
It has the values $\nu = ++$ for 
$S^{(1)}=S^{(2)}=1$, $+-$ for $S^{(1)}=-S^{(2)}=1$, $-+$ 
for $-S^{(1)}=S^{(2)}=1$  and $--$ for $S^{(1)}=S^{(2)}=-1$.
A complete configuration of the two Ising models is thus described
by the set $\{\nu_1, \ldots, \nu_N\}$. 

Since we are
interested in the time evolution not for a single sequence of $\xi_i(t)$,
but in $\xi$-averaged quantities we consider a whole ensemble of 
system pairs $(H^{(1)}$,$H^{(2)})$ and define a probability distribution 
\begin{equation}
P(\nu_1,\ldots,\nu_N,t) = \left \langle \sum_{\nu_i(t)} \prod_i 
\delta_{\nu_i,\nu_i(t)} \right \rangle
\end{equation}
where $\langle \cdot \rangle$ denotes the ensemble average.

\subsection{Transition probabilities}
In order to formulate a  master equation for the probability distribution
$P(\nu_1,\ldots,\nu_N,t)$ we need to know
the transition probabilities $w(\nu \rightarrow \mu)$
between
the states $\nu$ of  a spin pair. Since the Glauber dynamics (2) involves only
a single lattice site within each time step, we have to consider transitions
between the states $\nu$ of a single site only. Let us look, e.g., at the transition of  site
$i$ from state $--$ to  $++$.  This corresponds to both $S^{(1)}$  and $S^{(2)}$
changing from $-1$ to $1$.  According to the Glauber dynamics (2) this requires
$v(h_i^{(1)})-\xi_i > 0$ and $v(h_i^{(2)})-\xi_i > 0$. Since $v(h)$ is a monotonous
function of $h$ both equations are simultaneously fulfilled for
$v[\min(h_i^{(1)}, h_i^{(2)})]-\xi_i > 0$. Because $\xi_i$ is a random number taken
from a uniform distribution between 0 and 1,  the transition probability is given by
\begin{equation}
w(-- \rightarrow ++) = v[\min(h^{(1)},h^{(2)})] .
\end{equation}

Analogously, for a transition from state $--$ to $+-$ the two inequalities
$v(h_i^{(1)})-\xi_i > 0$ and $v(h_i^{(2)})-\xi_i < 0$ have to be fulfilled.
Since $v(h)$ is a monotonous function of $h$ this is only possible for 
$h_i^{(1)} > h_i^{(2)}$. The transition probability is obviously given by
\begin{equation}
w(-- \rightarrow +-) = \Theta(h^{(1)} - h^{(2)}) \left [ v(h^{(1)}) -v(h^{(2)}) \right ].
\end{equation}

The transition probabilities $w(\nu \to \mu)$ fulfill the following symmetry relations
\begin{mathletters}
\begin{eqnarray}
w(++ \rightarrow \nu) = w(-- \rightarrow \nu) \\ 
w(+- \rightarrow \nu) = w(-+ \rightarrow \nu)
\end{eqnarray}
\end{mathletters}
for any state $\nu$ as can easily be seen by making the substitutions
$S_i^{(n)}(t) \rightarrow -S_i^{(n)}(t)$ and $\xi_i(t) \rightarrow 1-\xi_i(t)$ 
on the right-hand side of (2).

The remaining transition probabilities can be calculated along the same lines 
as above. They are summarized in Table I.
\begin{table}
\caption{Transition probabilities $w(\nu \rightarrow \mu)$ between the states of
a spin pair}
\begin{tabular}{cc}
$-- \rightarrow --$ & $v[-\max(h^{(1)},h^{(2)})]$ \\
$-- \rightarrow ++$& $v[\min(h^{(1)},h^{(2)})]$ \\
$-- \rightarrow -+$ & $\Theta(h^{(2)} - h^{(1)}) \left [ v(h^{(2)}) -v(h^{(1)}) \right ]$ \\
$-- \rightarrow +-$ & $\Theta(h^{(1)} - h^{(2)}) \left [ v(h^{(1)}) -v(h^{(2)}) \right ]$ \\[1mm]
\hline\\[-2mm]
$-+ \rightarrow --$ & $\Theta(-h^{(1)} - h^{(2)}) \left [ v(-h^{(1)}) -v(h^{(2)}) \right ]$ \\
$-+ \rightarrow ++$ & $\Theta(h^{(1)} + h^{(2)}) \left [ v(h^{(2)}) -v(-h^{(1)}) \right ]$ \\
$-+ \rightarrow -+$ & $v[\min(-h^{(1)},h^{(2)})]$ \\
$-+ \rightarrow +-$ &$v[-\max(-h^{(1)},h^{(2)})]$ \\
\end{tabular}
\end{table}

\subsection{The master equation}
Having calculated the transition probabilities between the states $\nu$ of a spin pair
we are now in the position to write down the equation of motion for the probalitity
distribution $P(\nu_1, ...\nu_N,t)$.  It has the form of a usual master equation
\begin{eqnarray}
\lefteqn{\frac d {dt} P(\nu_1,\ldots,\nu_N,t) =} \nonumber\\
& &- \sum_{i=1}^N \sum_{\mu_i \not= \nu_i}
P(\nu_1,\ldots,\nu_i,\ldots,\nu_N,t) w(\nu_i \to \mu_i) \nonumber\\
& &+ \sum_{i=1}^N \sum_{\mu_i \not= \nu_i}
P(\nu_1,\ldots,\mu_i,\ldots,\nu_N,t) w(\mu_i \to \nu_i)
\end{eqnarray}
where the term in the second line describes the decrease of  
$P(\nu_1, ...\nu_N,t)$  due to the initial
configuration $\{\nu_1, \ldots \nu_N\}$ being changed at one of the sites $i$ from 
$\nu_i$ to $\mu_i$. The term in the third line of the master equation describes
the increase of $P(\nu_1, ...\nu_N,t)$ due to "scattering" from all the other states
into  $\{\nu_1, \ldots \nu_N\}$. Note that we have suppressed the factor $1/N$ in 
the transition probabilities which corresponds to random selection of one of the 
lattice sites in every time step. This neglect corresponds to a redefinition of the 
time scale (which is now independent of the system size)
and does not change the dynamic behavior. 

This master equation contains, of course, the full difficulty of the dynamic 
many-body problem. A complete solution is therefore out of question, and 
one has to resort to approximation methods. In the following section
we discuss how to construct a mean-field like approximation to the 
master equation (9).

\section{Effective-field approximation}
Usually a mean-field theory of a phase transition 
can be obtained by taking the range of the interaction to infinity:
\begin{equation}
J_{ij} = J_0/N \qquad   {\rm for \enspace all } \quad i,j
\end{equation}
In the thermodynamic limit $N \rightarrow \infty$ this suppresses all fluctuations. 
In particular,
the local magnetic fields $h_i^{(n)}$ of all sites in one system become equal
and identical to the mean-field value $J_0 m$. Since the two Ising models 
$H^{(1)}$ and $H^{(2)}$ are thermodynamically identical this leads to
$h_i^{(1)}=h_i^{(2)}$. However, some of the transition probabilities depend 
on the existence of fluctuations (see table I), i.e.
$w(\nu \rightarrow \mu)$ go to zero with $h^{(1)}-h^{(2)} \rightarrow 0$.
In particular, this is true for $w(-- \rightarrow -+)$ and $w(-- \rightarrow +-)$ which 
are responsible for increasing the damage $D$. Consequently, if the thermodynamic
limit and the limit of infinite range of the interaction are taken at a too early stage
of the calculation, the resulting model does not show any spreading of the damage.
To circumvent these problems we develop a slightly more sophisticated 
effective-field approximation that retains the fluctuations though in a quite
simplistic manner.  As will be shown in section IIIC, taking the range of  the interaction
to infinity within the framework of this approximation yields a sensible limit.

\subsection{Effective-field theory for short-range models}
The central idea of this effective-field method is to retain the fluctuations 
but to treat the fluctuations at different sites as
statistically independent. This amounts to approximating the 
probability distribution $P(\nu_1, \ldots, \nu_N,t)$ by a product of 
identical single-site distributions $P_{\nu}$,
\begin{equation}
P(\nu_1,\ldots,\nu_N,t) = \prod_{i=1}^N P_{\nu_i}(t).
\end{equation}
Inserting this into the master equation (9) and summing over all states of sites 
$i=2 \ldots N$ gives an equation of motion for
the single-site distribution $P_{\nu_1}$, 
\begin{equation}
\frac d {dt} P_{\nu_1} =  \sum_{\mu_1 \not= \nu_1} [- P_{\nu_1} W(\nu_1 \to \mu_1)
+ P_{\mu_1} W(\mu_1 \to \nu_1)],
\end{equation}
where 
\begin{equation}
W(\mu_1 \to \nu_1) = \left \langle w(\mu_1 \to \nu_1) \right \rangle_P
\end{equation}
is the transition probability averaged over the states $\nu_i$ of all sites 
$i \not= 1$ according to the distribution $P_{\nu_i}$. 
Since all sites of the systems are equivalent the site index  
$i$ will be suppressed from now on.

Note that the average magnetizations $m^{(1)}$, $m^{(2)}$
of the two systems and the Hamming distance $D$ can be
easily expressed in terms of $P_\nu$,
\begin{mathletters}
\begin{eqnarray}
m^{(1)} &=& P_{++} + P_{+-} - P_{-+} - P_{--},\\
m^{(2)} &=& P_{++} - P_{+-} + P_{-+} - P_{--}, \\
D &=& P_{+-} + P_{-+}.
\end{eqnarray}
\end{mathletters}

So far the considerations have been rather general,  in the following 
subsections we will apply the general formalism to different physical
situations. In section IIIB we investigate a two-dimensional system 
with short-range interactions and vanishing external field. We determine 
not only the location of the spreading transition but also calculate
the stationary states of the systems. Section IIIC
deals with the limit of infinite-range interactions, and in sec. IIID we 
study the influence of an external magnetic field on the spreading transition.

\subsection{Solution of a two-dimensional model}
In this subsection we investigate the damage spreading for a two-dimensional
Glauber Ising model on a hexagonal lattice (with each site having
three nearest neighbors). The interaction is taken to be 
a nearest-neighbor interaction of strength $J$ and the external magnetic field
is set to zero.

In order to solve the master equation (12) for the single-site distribution
$P_\nu$ we first determine the effective transition probabilities $W(\nu 
\rightarrow \mu)$. Let us calculate the
probabilities for a particular site $i$. 
To this end we have to average the transition probabilities
given in table I with respect  to the states of all other sites
of the system. However, since the interaction is between nearest
neighbors,  the transition probabilities depend on 
the states of these three neighbors of site $i$ only.
Each of the neighbors can be in one of  four states, thus we have to consider
64 different configurations of the neighboring sites.  The probabilities for 
these configurations and the resulting local magnetic fields are given in table II.
\begin{table}
\caption{Probabilities for the states of the three neighboring sites and resulting
local magnetic fields $h^{(1)}$ and $h^{(2)}$ 
for the two-dimensional case on the hexagonal lattice}
\begin{tabular}{ccccc}
& \multicolumn{4}{c} {$h^{(1)}$}  \\
$h^{(2)}$ & $3J$ & $J$ & $-J$ & $-3J$ \\
\hline\\[-2mm]
$3J$ & $P_{++}^3$ & $3P_{++}^2 P_{-+}$ & $3P_{++} P_{-+}^2$ & $P_{-+}^3$\\[2mm]
$J$ & $3P_{++}^2P_{+-}$ & $3P_{++}^2 P_{--} \ +$ & $3P_{+-} P_{-+}^2 \ +$ & $3P_{-+} P_{--}^2$ \\
& & $6P_{++} P_{+-} P_{-+}$ & $6P_{+-} P_{-+} P_{--}$ & \\[2mm]
$-J$ & $3P_{++} P_{+-}^2$ & $3P_{+-}^2 P_{-+} \ +$ & $3P_{++} P_{--}^2 \ +$ & $3 P_{-+} P_{--}^2$ \\
& &  $6P_{++} P_{+-} P_{--}$ & $6P_{+-} P_{-+} P_{--}$  &\\[2mm]
$-3J$ & $P_{+-}^3$ & $3P_{+-}^2 P_{--}$ & $3 P_{+-} P_{--}^2$ & $P_{--}^3$ \\
\end{tabular}
\end{table}

With the help of tables I and II the averaged transition probabilities
(13) can be easily calculated by adding up the
contributions of all 64 configurations. The resulting expression are quite lengthy
though simple. Therefore we present only the example
\begin{eqnarray}
&W&(-- \rightarrow ++) =\nonumber \\
& &\langle w(-- \rightarrow ++) \rangle = \langle v[\min(h^{(1)},h^{(2)}] \rangle = \nonumber \\
& &P_{++}^3 v_3 + 3P_{++}^2 P_{-+} v_1 + 3P_{++} P_{-+}^2 v_{-1} + P_{-+}^3 v_{-3}+ \nonumber\\
& &(3P_{++}^2P_{+-} + 3P_{++}^2 P_{--}  + 6P_{++} P_{+-} P_{-+})  v_1 + \nonumber\\
& &(3P_{+-} P_{-+}^2  + 6P_{+-} P_{-+} P_{--}) v_{-1} + 3P_{-+} P_{--}^2 v_{-3} + \nonumber \\
& &(3P_{++} P_{+-}^2 +3P_{+-}^2 P_{-+}  + 6P_{++} P_{+-} P_{--}) v_{-1} + \nonumber\\
& &(3P_{++} P_{--}^2 + 6 P_{+-} P_{-+} P_{--}) v_{-1} + 3 P_{-+} P_{--}^2 v_{-3} + \nonumber\\
& &(P_{+-}^3  + 3P_{+-}^2 P_{--} + 3 P_{+-} P_{--}^2 + P_{--}^3) v_{-3}
\end{eqnarray}
with
\begin{equation}
v_n = v(nJ).
\end{equation}

Equations of motion for the magnetizations $m^{(1)}$ and $m^{(2)}$ as well as
for the damage $D$ can be derived by inserting the definitions (14) into the 
single-site master equation (12).  After some manipulations the
equations of motion for the magnetizations read
\begin{eqnarray}
\frac d {dt} m^{(n)} &=& m^{(n)} \left\{ -1 + \frac 3 4 [\tanh(3J/T) +\tanh(J/T)]   \right\} 
\nonumber \\ &+& \frac 1 4 (m^{(n)})^3 \left[ \tanh(3J/T) - 3\tanh(J/T) \right ].
\end{eqnarray}
These equations are, of course, identical to the equation of motion of the magnetization
derived for a single system within the same framework of statistically 
independent fluctuations.  The point at which the coefficient of the
term linear in $m$ on the right-hand side of (17) changes sign defines the
(equilibrium) critical temperature $T_C$ of the Ising model within our approximation.
$T_C$ is thus determined by
\begin{equation}
\frac 3 4 [ \tanh (3J/T_C) + \tanh(J/T_C)]  = 1
\end{equation}
which gives $T_C/J \approx 2.104$. 
The stationary solution of  (17) can be used to determine the magnetization
as a function of temperature. For temperatures $T<T_C$ we obtain
\begin{equation}
(m^{(n)})^2 = \frac {\frac 3 4 [\tanh(3J/T) + \tanh(J/T)]-1} 
 {\frac  3 4 \tanh(J/T) - \frac 1 4 \tanh(3J/T)}.
\end{equation}

We now turn to the discussion of the Hamming distance $D$.
After inserting (14) into (12) the equation of motion of the Hamming 
distance $D$  can be written as
\begin{eqnarray}
\frac d {dt} D&=&(1-D) [ W(-- \rightarrow +-) + W(-- \rightarrow -+)] +\nonumber \\
&+& D [-1 + W(-+ \rightarrow +-) + W(-+ \rightarrow -+)].
\end{eqnarray}
Since in the following we will be mainly interested in the stationary solutions 
of this equation we restrict the considerations to cases where both
systems are in equilibrium at the beginning of the damage spreading 
process. In doing so we exclude, however, all phenomena connected
with the behavior after a quench from high temperatures to temperatures
below $T_C$. These phenomena require an investigation of the {\em early}
time behavior and will be analyzed elsewhere \cite{vojta2}. 

It is now useful to distinguish three cases, (i) damage spreading in the 
paramagnetic phase ($T>T_C$), (ii) the ferromagnetic phase ($T<T_C$) 
where both systems are in the same pure state (i.e., free energy valley), 
$m^{(1)} = m^{(2)} =m$ and (iii)  the ferromagnetic phase ($T<T_C$) 
where the two systems are in different pure states,
$m^{(1)} = -m^{(2)} =m$.

\subsubsection{Paramagnetic phase}
In the paramagnetic phase all $P_\nu$ can be expressed in terms of  $D$:
\begin{mathletters}
\begin{eqnarray}
P_{++} = P_{--}&=& \frac 1 2 (1-D), \\
P_{+-} = P_{-+} &=& \frac 1 2 D.
\end{eqnarray}
\end{mathletters}
By inserting this into the transition probabilities $W(\nu \rightarrow \mu)$
calculated from (13) and table II the equation of motion (20) of the Hamming
distance $D$  can be written as
\begin{equation}
\frac d {dt} D = \frac 1 2 (D -3 D^2 +2 D^3) \tanh (3J/T) .
\end{equation}
This equation has three stationary solutions (fixed points), viz.
$D_1^*=0$ which corresponds to both systems being identical,
$D_2^*=1$ where $S^{(1)} = - S^{(2)}$ for all sites
and $D_3^*=1/2$ which corresponds to the two systems being 
completely uncorrelated \cite{glausym}.
To determine the stability of the fixed points we linearize the equation
of motion (22) in $d_k=D-D_k^*$.  The linearized equation has the solution
\begin{equation}
d_k(t) = d_{k0} \ e^{\lambda_k t}
\end{equation}
with $\lambda_1 = \lambda_2 = \frac 1 2 \tanh(3J/T)$ and 
$\lambda_3 = - \frac 1 4 \tanh (3J/T)$.
Consequently, the only stable fixed point is $D_3^*= 1/2$. In the
whole paramagnetic phase the damage spreads and asymptotically
reaches the value $D=1/2$.
If the two systems start very close together ($D$ small initially)
their distance in phase space increases exponentially with a
Lyapunov exponent $\lambda_1  = \frac 1 2 \tanh(3J/T)$. Therefore
the Glauber dynamics shows chaotic behavior in the whole 
paramagnetic phase. Note, that for large temperatures
the Lyapunov exponent $\lambda_1$ goes to zero as 
$\lambda_1 \sim 3J/T$.  Thus, the time it takes the system to
reach the stationary state $D= D_3^* =1/2$ diverges for
$T \rightarrow \infty$.  This has recently also been found in 
simulations \cite{wappler}. The dependence of the 
Lyapunov exponent on temperature is presented in fig. 1.
\begin{figure}
  \epsfxsize=7.0cm
  \epsfysize=7.0cm
  \centerline{\epsffile{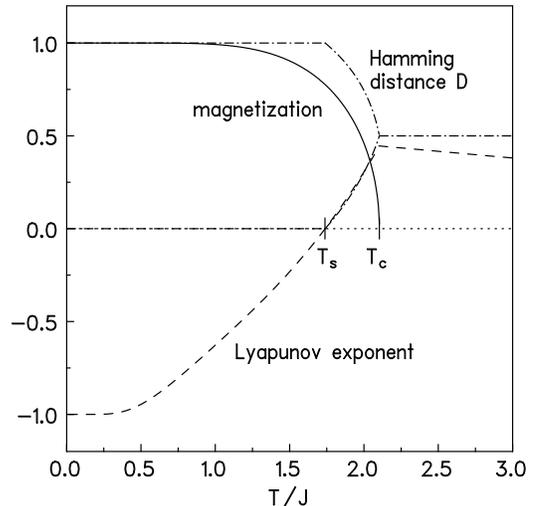}}
  \caption{Magnetization $m$, asymptotic Hamming distance $D^*$ and Lyapunov
   exponent $\lambda_1$ as functions of temperature for the Glauber Ising
   model. Below $T_c$ the curve for $D^*$ has two branches corresponding to
   the two systems being in the same or in different free energy valleys.}
  \label{fig:1}
\end{figure}

\subsubsection{Ferromagnetic phase with $m^{(1)} = m^{(2)} =m$}
In this paragraph we study the case where both systems are in 
the same free energy valley.
The single-site probabilities $P_\nu$ can be expressed in terms of 
$D$ and $m$:
\begin{mathletters}
\begin{eqnarray}
P_{++} &=& \frac 1 2 (1 -D +m), \\
P_{--} &=& \frac 1 2 (1-D -m), \\
P_{+-} = P_{-+} &=& \frac 1 2 D.
\end{eqnarray}
\end{mathletters}
After inserting this into the averaged transition probabilities (13) the equation
of motion of the Hamming distance takes the form
\begin{eqnarray}
\frac d {dt} D &=& \frac 1 2 (D - 3 D^2 + 2 D^3) \tanh (3J/T)  \nonumber \\
& -& \frac 3 4 m^2 [2 D \tanh (J/ T)\nonumber \\
& & \qquad -D^2 \tanh (J/ T) +D^2 \tanh (3J/T) ].
\end{eqnarray}
This equation has two fixed points $D^*$ in the interval [0,1]. The first
fixed point is $D_1^*=0$. By linearizing (25) in $d_1=D-D_1^*$ we investigate 
the stability of this fixed point. We again find that $d_1(t)$ follows the 
exponential law  (23) with $\lambda_1= \frac 12 \tanh (3J/T)$
$- \frac 3 2 m^2 \tanh (J/T)$. Using for $m^2$ the expression (19) it is easy
to discuss the behavior of $\lambda_1$. For temperatures
larger than a spreading temperature $T_S$ which is defined by
\begin{equation}
3 m^2 \tanh (J /T_S) =\tanh (3J/T_S)
\end{equation}
the Lyapunov exponent $\lambda_1$ is positive and 
thus the fixed point $D_1^*$ is unstable.
For $T<T_S$ the Lyapunov exponent $\lambda_1$ is negative and the
fixed point $D_1^*$ is stable. Consequently, the Glauber dynamics is chaotic
for temperatures above $T_S$ but regular below.
Eq. (26) gives $T_S \approx 1.739 J \approx 0.826 T_C$. 

For temperatures $T>T_S$ the equation of motion (25)
possesses another fixed point $D_3^*$ with $0< D_3^* <1/2$ 
which is always stable. Its temperature dependence is presented
in fig. 1. Close to the spreading temperature the
asymptotic Hamming distance $D_3^*$ increases linearly with $T-T_s$ which
corresponds to the spreading transition being of 2nd order.
The order parameter exponent $\beta$, defined by $D^*=|T-T_s|^\beta$
is given by $\beta=$.
In contrast to the paramagnetic phase, where the two systems eventually 
become completely uncorrelated, for $T_s<T<T_c$ the asymptotic
Hamming distance $D$ is always smaller than 1/2 so that the two systems
remain partially correlated (as it must be the case
since both systems are in the same free energy valley).
Directly at the spreading point the term linear in $D$ in (25) vanishes.
For small Hamming distances the equation of motion now reads
$d D /dt \propto -D^2$ which 
gives a power-law behavior $D(t) \propto t^{-\delta}$. The
critical exponent is given by $\delta=1$.

\subsubsection{Ferromagnetic phase with $m^{(1)} = -m^{(2)}=m$}
We now turn to the case where the two systems are in different
free energy valleys. The single-site probabilities $P_\nu$ 
can be expressed in terms of  $D$ and $m$:
\begin{mathletters}
\begin{eqnarray}
P_{++} =P_{--} &=& \frac 1 2 (1 -D), \\
P_{+-} &=& \frac 1 2 (D+m) , \\
P_{-+} &=& \frac 1 2 (D-m).
\end{eqnarray}
\end{mathletters}
With this substitutions the equation of motion (20) of the Hamming distance
can be written as
\begin{eqnarray}
\frac d {dt} D &=& \frac 1 2 (D - 3 D^2 + 2 D^3) \tanh (3J/T)  \nonumber \\
& +& \frac 3 4 m^2 [ \tanh(3J/T) +\tanh(J/T) - 2 D \tanh (3J/ T)\nonumber \\
& & \qquad +D^2 \tanh (3J/ T) -D^2 \tanh (J/T) ].
\end{eqnarray}
Analogously to the preceeding paragraph, this equation possesses two
fixed points.  The fixed point $D_2^*=1$ exits for all temperatures. It is stable
for temperatures below $T_S$ and unstable above. For $T>T_S$  
(28) has  another fixed point , $D_4^*$ with $1/2 <D_4^* <1$ which is 
always stable. Its temperature dependence is given in fig. 1.

\subsection{The limit of high dimensions}
In this subsection we study damage spreading in the Glauber Ising
model in the limit of high dimensions, i.e. in the
mean-field limit proper. Within the framework of
our effective field approach high dimensions correspond to
high coordination numbers, i.e. high numbers of nearest neighbors. 
We therefore consider  a Glauber Ising model on a lattice with
$z$ nearest neighbors and study the limit $z \rightarrow \infty$.
To obtain a physically sensible limit we scale the interaction strength
with $z$, $J=J_0/z$. 

In the limit $z \to \infty$ the thermodynamics is described by the usual
mean-field theory. The equilibrium critical temperature is given by
$T_C=J_0$ and in the ferromagnetic phase the magnetization is 
determined by the equation of state
\begin{equation}
m = \tanh(mJ_0/T).
\end{equation}

In order to determine the spreading temperature $T_S$ it is sufficient to 
study the equation of motion (20) of the Hamming distance to linear order 
in $D$. To this end we have to determine $W(-+ \to +-)$ and $W(-+ \to -+)$
to zeroth order in $D$ but $W(-- \to +-)$ and $W(-- \to -+)$
to linear order in $D$.

To zeroth order in $D$ we have $h^{(1)}=h^{(2)} =h$ and
in the limit $z \to \infty$ $h$ is $\delta$-distributed at $h=J_0 m$.
The transition probabilities are thus given by (see table I)
\begin{mathletters}
\begin{eqnarray}
W(-+ \to +-) &=& v[-\max(-h^{(1)},h^{(2)})] \nonumber \\
 &=& v(-J_0|m|), \\
W(-+ \to -+) &=& v[\min(-h^{(1)},h^{(2)})]  \nonumber \\
 &=& v(-J_0|m|).
\end{eqnarray}
\end{mathletters}

We now calculate $W(-- \to +-)$ and $W(-- \to -+)$
to linear order in $D$. These transition probabilities do 
not have a zeroth-order contribution.  In linear order in $D$
only those configurations of the $z$ neighboring sites  
contribute for which  the two  systems differ in the state of 
a single site.
In this case $h^{(1)}$ and $h^{(2)}$ differ by $2J_0/z$.
Therefore we obtain
\begin{mathletters}
\begin{eqnarray}
W(-- \to +-)&=& \Theta(h^{(1)} - h^{(2)}) [v(h^{(1)} ) - v(h^{(2)})] \nonumber\\
&=& z \ P_{+-}  (2J_0 /z) \  v'(J_0m).
\end{eqnarray}
Here $v'(h)$ is the derivative of $v$ with respect to its argument. The
additional factor $z$ in the second line arises since each of the $z$
neighbors can be the one where the two systems differ. Analogously we obtain
\begin{eqnarray}
W(-- \to +-)&=& \Theta(h^{(2)} - h^{(1)}) [v(h^{(2)} ) - v(h^{(1)})] \nonumber\\
&=& z \ P_{-+}  (2J_0 /z) \  v'(J_0m).
\end{eqnarray}
\end{mathletters}
By inserting these results for the transition probabilities into the
equation of motion (20) of the Hamming distance we find
\begin{equation}
\frac d {dt} D =\lambda D 
\end{equation}
where the Lyapunov exponent $\lambda$ is given by
\begin{eqnarray}
\lambda &= & -1 + 2 \frac {\exp(-J_0|m|/T)} {\exp(J_0|m|/T) 
 +\exp(-J_0|m|/T)} 
\nonumber \\
 & &+  \frac {4J_0/T} {[\exp(J_0m/T) +\exp(-J_0m/T)]^2} .
\end{eqnarray}
This can be simplified to
\begin{equation}
\lambda = -m +(1-m^2) J_0/T.
\end{equation}
In the paramagnetic phase ($m=0$) the Lyapunov exponent is simply
$\lambda=J_0/T>0$. Thus the Glauber dynamics is chaotic in the whole 
paramagnetic phase.

The temperature dependence of the Lyapunov exponent 
$\lambda$ in the 
ferromagnetic phase is presented in fig. 2.
\begin{figure}
  \epsfxsize=7.0cm
  \epsfysize=7.0cm
  \centerline{\epsffile{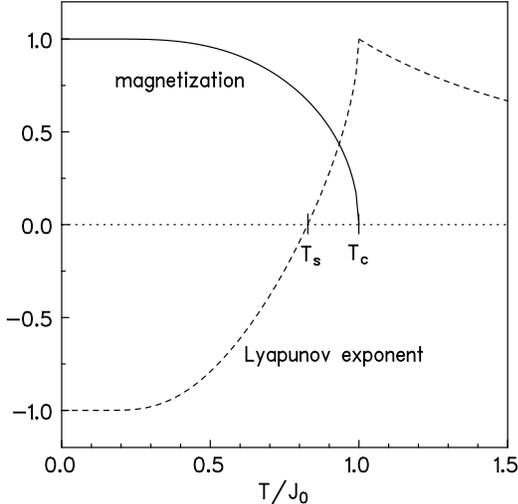}}
  \caption{Magnetization $m$ and Lyapunov
   exponent $\lambda_1$ as functions of temperature for the Glauber Ising
   model with vanishing external field in the limit of high dimensions.} 
  \label{fig:2}
\end{figure}
$\lambda$ changes sign at $T_S \approx 0.827 J_0 = 0.827 T_C$. 
Consequently, the dynamics is chaotic for temperatures larger than 
$T_S=0.827 J$ and regular for temperatures smaller than $T_S$.
Note that the value for $T_S/T_C$ for the two-dimensional model
of  sec. IIIB is very close to but not identical to the value for the case 
$z \to \infty$.

\subsection{Damage spreading in a field}
In this subsection we generalize the effective-field theory  to a finite 
external magnetic field $h$.  For simplicity, we do this only for the 
model introduced in the preceeding subsection, viz. the limiting case 
of high dimensions.

The equation of state (29) has to be replaced by
\begin{equation}
m = \tanh[(mJ_0+h)/T].
\end{equation}
Analogously, in all transition probabilities $W(\nu \to \mu)$ the term
$J_0m$ has to be replaced by $J_0m+h$.
After inserting the transition probabilities into the equation of motion (20)
of the Hamming distance one finds $d/ dt \ D = \lambda D$ and the
Lyapunov exponent can again be expressed in terms of the magnetization:
\begin{equation}
\lambda = -m +(1-m^2) J_0/T.
\end{equation}
The temperature and field dependence of the Lyapunov exponent
is illustrated in fig. 3.
\begin{figure}
  \epsfxsize=7.4cm
  \epsfysize=7.0cm
  \centerline{\epsffile{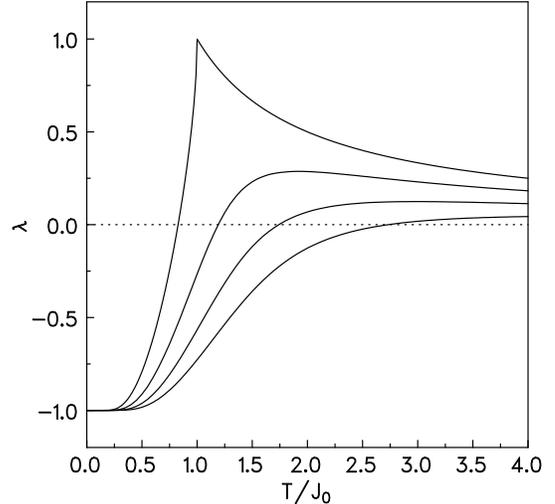}}
  \caption{Lyapunov
   exponent $\lambda$ as a function of temperature 
   for external fields $h=0, 0.2, 0.4$ and $0.6$ 
   (from up to down).} 
  \label{fig:3}
\end{figure}
Obviously, the external field shifts the spreading temperature
to higher values, thus suppressing chaotic behavior
and stabilizing the regular phase.
The phase boundary between the chaotic and the 
regular phase can be easily 
determined by solving the equation $\lambda=0$.  The resulting phase
diagram is presented in fig. 4. For comparison we also
give simulation results \cite{lecear} for a three-dimensional
Glauber Ising model.  
\begin{figure}
  \epsfxsize=7.4cm
  \epsfysize=7.0cm
  \centerline{\epsffile{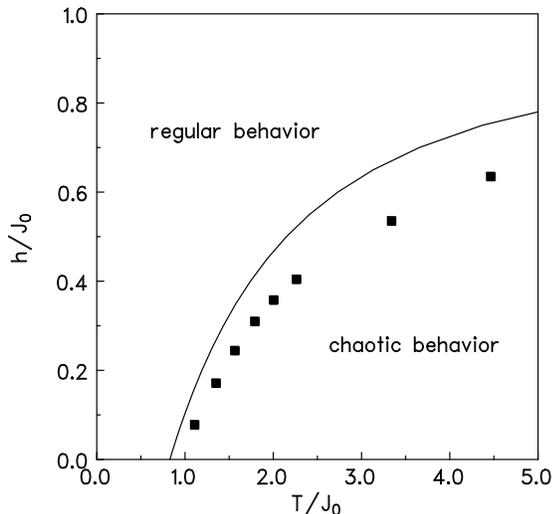}}
  \caption{Phase diagram of  damage spreading in the Glauber
    Ising model in the limit of infinite dimensions. The full line 
    shows the result of our theory, the squares are simulation
    results of Le Ca\"er with $T$ and $h$ rescaled by $T_C$. } 
  \label{fig:4}
\end{figure}
An investigation of  (35) and (36) for large temperatures shows that
the spreading temperature $T_S(h)$ diverges for $h/J_0 \to 1$
as
\begin{equation}
T_S(h)/J_0 = 1/(1-h/J_0) \qquad {\rm for} \ h/J_0 \to 1.
\end{equation}
Consequently, for external fields $h>J_0$ the dynamics is always 
regular.

\section{Conclusions}
To summarize, we have developed a master equation approach to
damage spreading and applied it to the Glauber Ising model. 
The master equation is an exact description of the damage
spreading problem, it does not contain any approximations.
We have then solved the master equation within an 
effective-field theory for
various physical situations.

In this final section we discuss some aspects which have not been 
covered yet. First , we compare the results of our effective-field theory
with numerical simulations of damage spreading of the Glauber
Ising model in two and three dimensions \cite{ts,grassberger,lecear}. 
In agreement with the simulation results we find a spreading transition
{\em below} the equilibrium critical temperature of the Ising model.
Our mean-field value $T_s/T_c \approx 0.827$ is considerably lower
than the latest numerical values \cite{grassberger} of 0.992 for a two-dimensional
and 0.922 for a three-dimensional Glauber Ising model. We expect
our value to be exact, however, for an infinite-dimensional model or,
equivalently, for infinite range of the interaction.
Grassberger \cite{conjecture} conjectured that the damage spreading 
transition in the Glauber Ising model is in the universality class
of directed percolation. Our results are compatible with that, since
the values of the critical exponents $\beta$ (which describes the
dependence of the stationary damage on the reduced temperature) 
and $\delta$ (which describes the time decay of the damage at the
spreading temperature) are identical to the mean-field values
$\beta = \delta =1$ of directed percolation.

Second, we want to clarify the relation to damage spreading in an
Ising model with {\em heat-bath} dynamics.  As already discussed in
the introduction the heat-bath Ising model does not show any
spreading of damage within a single pure state (free energy valley).
When applying our effective-field theory to the heat-bath Ising model
we find \cite{vojta} only a single fixed point $D_1^*=0$ if both 
systems are in the same pure state \cite{sym}. If the two systems are in
different pure states (for $T<T_C$)  we also find a single
fixed point only, viz $D_2^*=m$. Thus there is no chaotic behavior
within one pure state. However, the damage can spread 
(or, at least, will not heal) in the
heat-bath Ising model if the two copies start in different pure states
or choose to develop into different pure states after a quench from
high temperatures. For this case a mean-field theory
similar to our's has been considered before \cite{derridamf}.

Finally, we discuss possible extensions of the present
theory. In principle, the master equation approach
of sec. II can be applied to any damage spreading
problem in which the dynamics of a single system
is given by a stochastic map as in (2) (or a more general
map that involves several sites in each time step).
It would be very interesting to remap the master equation
onto a field theory and then apply renormalization group
methods to determine the critical behavior.

An obvious idea is to include quenched disorder
into the Hamiltonian of the Ising model either
in the form of a random external field or in the form of random 
interactions. Such systems have been numerically investigated in some 
detail, in particular in the case of random interactions 
\cite{spinglass}. Recently some interesting results 
have also been achieved for random fields \cite{wappler}.
Some investigations on the application of the
master equation approach
to disordered systems are in progress.

\acknowledgements
This work was supported in part by the DFG under grant number Vo 659/1-1
and by the NSF under grant number DMR-95-10185.


\begin{references}
\bibitem{glausym} For vanishing external field the Glauber algorithm
is invariant under a global flip of all spins. Thus, the existence of $D_1^*=0$
automatically implies the existence another fixed point, $D_2^*=1$, with  
the same stability properties.
\bibitem{sym} In contrast to the Glauber dynamics the heat-bath algorithm
does not preserve the symmetry with respect to a global flip of all spins.
Therefore $D=1$ is not a fixed point here.
\end{references}
\end{document}